\documentclass[a4paper,conference]{IEEEtran}
\IEEEoverridecommandlockouts
\usepackage{float}
\usepackage{wrapfig}
\usepackage[table,xcdraw]{xcolor}
\usepackage{wrapfig}
\usepackage{subcaption}
\usepackage{caption}
\usepackage{hyperref}
\def\etal{\emph{et al.}}

\ifCLASSINFOpdf
  \usepackage[pdftex]{graphicx}
  \graphicspath{{images/}}
\else
\fi
\begin{document}
%
\title{Can viewer proximity be a behavioural marker for Autism Spectrum Disorder?}

\author{
  \IEEEauthorblockN{R. Bishain\IEEEauthorrefmark{1},
    B. Chakrabarti\IEEEauthorrefmark{2},
    J. Dasgupta\IEEEauthorrefmark{4},
    I. Dubey\IEEEauthorrefmark{2}\IEEEauthorrefmark{3},
    Sharat Chandran\IEEEauthorrefmark{1}
    }
\IEEEauthorblockA{[On behalf of the START consortium.]}
  \IEEEauthorblockA{\IEEEauthorrefmark{1}
    Department of Computer Science and Engineering, Indian Institute of Technology Bombay, India}
  \IEEEauthorblockA{\IEEEauthorrefmark{2}
    Centre for Autism, School of Psychology \& Clinical Language Sciences, University of Reading, UK} 
  \IEEEauthorblockA{\IEEEauthorrefmark{3}
    Division of Psychology, De Montfort University, UK}
  \IEEEauthorblockA{\IEEEauthorrefmark{4}
    Child Development Group, Sangath, Bardez, Goa, India}
  }

\maketitle


\begin{abstract}
  Screening for any of the Autism Spectrum Disorders is a complicated
  process often involving a hybrid of behavioural observations and
  questionnaire based tests. Typically carried out in a controlled
  setting, this process requires trained clinicians or psychiatrists
  for such assessments.  Riding on the wave of technical advancement
  in mobile platforms, several attempts have been made at
  incorporating such assessments on mobile and tablet devices.  
  
In this paper we analyse videos generated using one such screening
  test.  This paper reports the first use of the efficacy of using the
  observer's distance from the display screen while administering a
  sensory sensitivity test as a behavioural marker for Autism for
  children aged 2-7 years.  The potential for using a test such as
  this in casual home settings is promising.
\end{abstract}


%
\IEEEpeerreviewmaketitle

\section{Introduction}
\label{sec:intro}



Autism Spectrum Disorders (ASD) are associated \cite{dsm5} with
atypical social communicative behaviour, restricted range of interests
and repetitive behaviour. While the global population-level prevalence
of ASD is estimated to be around 1\%, the challenges faced in low and
middle income countries (LMIC) are significantly greater than those
faced by high income countries (HIC) due to the shortage of resources
and expertise. Assessment of ASD constitutes the first hurdle in
meeting this important global health need.

\begin{figure*}
\centering
\includegraphics[width=0.9\textwidth]{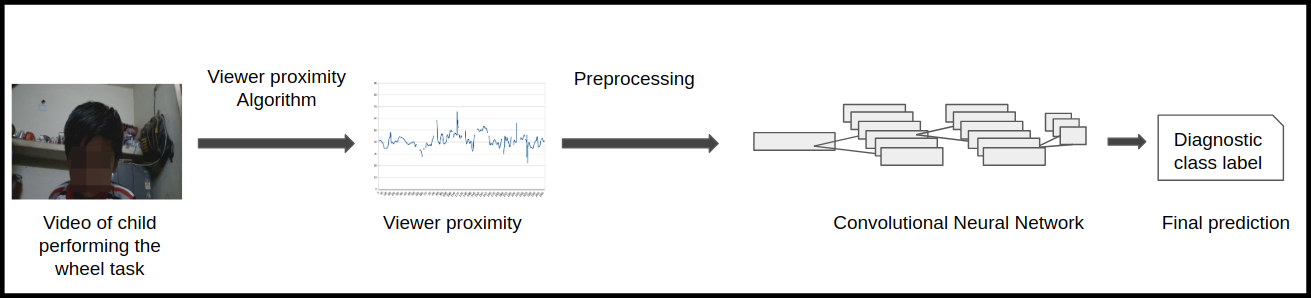}
\caption[Process pipeline]{Overview: The contribution in this paper is to provide (and validate) a method of obtaining viewer distances for children whose pictures are taken in a casual home setup with
  uncalibrated low resolution cameras and utilise it for diagnostic classification of autism.}
\label{fig:Process}
\end{figure*}

Patients going to the hospital, even assuming the availability of
trained professionals is not scalable. In 2018, Egger \etal \cite{egger2018} reported the use of an iPad based method to be used
in a home setting in HIC; from the computer vision perspective, the
technology of interest was to automatically obtain head pose based on
the method in \cite{intraface} and emotion and attention analysis
\cite{Hashemi}.

These autism screening and diagnostic analyses typically focus on
social functioning measures. However, many autistic individuals show
an atypical \emph{sensory sensitivity} in one or more modalities
\cite{bhisma}. Indeed, recognition of this gap in assessments led to
sensory symptoms being included within the diagnostic framework for
ASD in the latest version \cite{dsm5} of the diagnostic manual. The
sensory sensitivity test in this paper follows ideas similar to
\cite{egger2018} but includes the
`wheel-task'(Sec.~\ref{sec:wheelTask}) and is administered in LMIC on casual Android platforms.

The problem solved in this paper is illustrated in
Figure~\ref{fig:Process} (refer next page). To the best of our knowledge, this is the
first evidence of using a computer vision algorithm for the wheel task
for a significant medical condition, with emphasis that the target
audience is children in a relaxed LMIC home setting.

\subsection{The wheel task}
\label{sec:wheelTask}

The objective of the wheel task is to present a `rotating' disk (refer below) on the tablet screen to a child who is being screened for ASD.Autistic children tend to look intently \cite{tavassoli} at a spinning wheel as well as come closer to the wheel. While the child is engaged in this task, the front camera of the device records the child's movement.


\begin{wrapfigure}{r}{0.3\textwidth}
\includegraphics[width=0.3\textwidth]{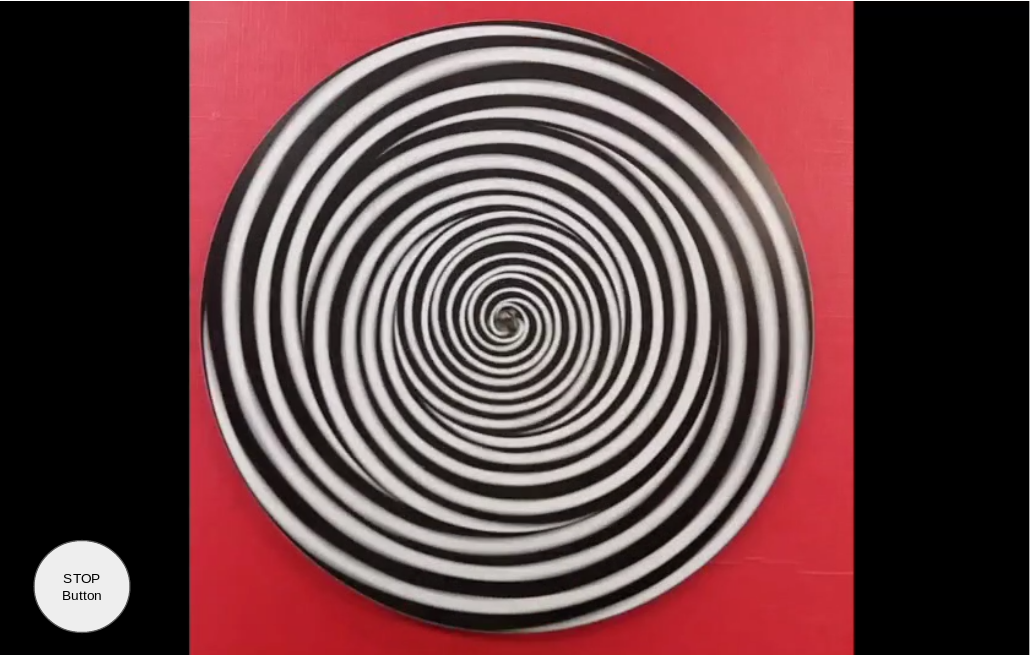}
A spinning wheel is displayed on a tablet screen. Size of the disk as well as the display screen are uniform for all subjects (children)
\label{Wheel}
\end{wrapfigure}

In the current work, we investigate the relative movement of the child
with respect to the screen.  If a child is fascinated by the spinning
object they tend to often move closer to the object and thus closer to
the front camera. In this paper, we posit that a measure of the
relative motion of the child with respect to the front camera can be
used as an indicator of their interest in the spinning object.  We
then utilise this observation for predicting the 
diagnostic classification of the child.


\textbf{Challenges:} Although cues such as relative face size can be
considered for computing the distance of the child from the screen, we
found in our experiments that these proxies for distance are
unreliable. The key computational challenge stems from our assumption
that different children could be seated at different distances --
indeed some children may even run the test in a supine position. Also,
the head orientation of a subject can inordinately impact the
perceived face size in a two-dimensional image. 

It is to be noted that while we administered the tests in an
unconstrained home environment, such tests are usually administered in
controlled laboratory settings. Also, the tests were conducted by
low-skilled health workers in LMIC households. The workers were not
provided any special training, unlike the clinicians who usually
conduct the tests in controlled settings. Hence, our approach
addresses this constrained context and aims to capture the viewer's
proximity using images captured from an uncalibrated low resolution
camera.  Please see the supplementary video on the data capture.

\subsection{Contributions}
Tablets recording viewers is becoming increasingly common, to say the
least, and face identification is standard.  The key technical
contributions in this work are

\begin{itemize}
\item A method that computes in metric units the distance between the
  front camera of a tablet and the face of a human being. This measure is often a requirement in such diagnostic analyses. (The source
  code will be provided)
\item A method that works
\begin{itemize}
    \item for \emph{children} as they tend to get distracted quickly and may not remain stationary in front of the tablet device during the task.
    \item in poorly lit conditions as the study is aimed at children from less privileged backgrounds. (Methods that work in well-lit homes may not generalise across low resolution, poorly lit environments. In this paper, we  demonstrate evidence that current computer vision methods are quite robust in home settings.)
\end{itemize}
\item Quantitative evidence that viewer proximity in the wheel task can be a marker for classification in child development.  (All earlier work on this task were subjective lab-based assessments. We report the first classifier which utilises this measure for diagnostic classification). 
\end{itemize}

The rest of this paper is organised as follows. We first discuss
related work, and follow it up with the method suggested and its
validation (Section~\ref{sec:method}). Results are presented in
Section~\ref{sec:results}.

\subsection{Related Work} Since ASD is a spectrum of developmental disorders, most of the diagnostic and screening assessments revolve around the analysis of behavioural symptoms. These can be analysed in dyadic interaction between the subject and others. For instance, in the case of young children, by observing the interaction between mother and child, psychologists can devise an appropriate intervention
plan (i.e.,  therapy) for better parental care and child development.

In such assessments \cite{112-NIH} the aim is to capture important
behavioural cues such as shared mutual attention, imitating other's
actions, etc. Such play interactions are also analysed for screening
and diagnostics \cite{103-ADOS} purposes and for longitudinal impact
assessments \cite{104-DCMA} of the selected intervention therapy.

Researchers have tried to employ machine vision techniques to automate
such analysis but they are either dependent on a multimodal setup or
analyse a very structured set of actions under a constrained
paradigm. For instance, Rehg \etal \cite{6619282} have analysed
behaviour of children by creating an independent Rapid-ABC paradigm. In
contrast, Marinoiu \etal \cite{110-Marinou} have attempted both action
and emotion recognition tasks in an unstaged environment. However,
they have focused on robot assisted therapy and not on the diagnostic
aspect of autism which is the focus of our paper.

The analysis of head movement, while children perform autism specific
assessments, has gained traction in the computer vision community in
recent years. More specifically, Martin \etal \cite{Martin2018}
contrasted head movements between ASD and non ASD groups of children
when watching videos of social vs non-social stimuli. They have
studied group differences vis-\`a-vis pitch, yaw and roll of the head
movements. Ogihara \etal \cite{Ogihara_2019_CVPR_Workshops} studied
specific temporal patterns of head movements to understand their
diagnostic potential.

The studies based on head movement \cite{Martin2018} and
\cite{Ogihara_2019_CVPR_Workshops}, discussed above, have been carried
out in laboratory settings. Children are seated in front of a monitor,
with a mounted camera, which records their movement while they view
various stimuli on the screen. \textit{In contrast, our work relies solely on
videos captured at the home of participant children.} The videos are
captured from the relatively weak front camera ($0.35$ Megapixels) of
Android tablet devices while the children view the stimuli presented
on the tablet screen.

\section{Methodology}
\label{sec:method}
The complete process pipeline, from data capture in the form of videos, to the final diagnostic class prediction, is depicted diagrammatically in Figure~\ref{fig:Process}. The details of each phase in the process pipeline are elaborated in Sections~\ref{2.1} to \ref{2.3}.

\subsection{Data Collection}\label{2.1}

As mentioned, several tasks are administered in determining markers
for autism. Typically, the battery of incorporated tasks are carried
out in controlled settings in a laboratory with expensive setups,
sensors and trackers. In contrast, we are limited in our approach due
to the requirements of a low-resource setting. Due to this requirement
the data collection was performed using an economical tablet device
with additional limitations of:
\begin{itemize}
\item No user dependence is permitted (For example, no task or user
  specific calibration was allowed).
\item No separate recording -- the recording was only via a low
  resolution RGB front camera that could be triggered with the task.
\item No additional user sensors or wearable setup was allowed.
\end{itemize}

\textbf{These limitations are to be contrasted with prior work}.  The tasks
were conducted by non-specialist community health workers on young
children at their homes. This was essential for the desired
scalability and for collecting data on autism related symptoms in
LMIC. In summary, the task was administered in as natural a setting as
possible.  In this paper, we have focused only on one of the several
tasks which have been incorporated for screening of autism risk.

\begin{table}[H]
\begin{center}
\begin{tabular}{|p{4.5cm}|c|c|}
\hline
Diagnostic Category & Video Count & Videos Used\\
\hline\hline
 (TD) Typically Developing & 39 & 37         \\
(ASD) Autism Spectrum Disorder     & 43  & 41        \\
(ID) Intellectually Disabled  & 39 & 37         \\
\hline
Total                    & 121 & 115  \\     
\hline
\end{tabular}
\end{center}
\caption[Wheel task video dataset]{Number of wheel task videos
  collected across the diagnostic categories of TD, ASD and ID
  children. All data collected on the same device model. Maximum video duration was 80 seconds.  As per our knowledge this is the largest dataset for such a task captured in home settings}
\label{table:Data Set Count}
\end{table}

The data is a set of videos of children performing the wheel task. Prior to the data collection the children were categorised in one of the three diagnostic groups by experts in a prestigious medical school-cum-hospital-cum-medical research university that sees a daily footfall of over thousand patients of all ages. In contrast, the video data set is collected by minimally trained workers who have administered these tasks in semi-urban households. A corpus of $121$ videos has been created (with $195000$ image frames). The videos have been captured across the diagnostic categories or classes of TD (Typically Developing), ASD and ID (Intellectual Disability) children as shown in Table~\ref{table:Data Set Count}.

These videos were manually verified for usability in our task and some
videos were observed to be unusable due to insufficient length of
relevant footage. Finally, after removing such videos a total of 115
videos were used for this analysis with the respective distribution
across the three diagnostic classes as shown in Table~\ref{table:Data
  Set Count}. We have included a sample video from this dataset in the
supplementary material.

 \subsection{Viewer proximity algorithm}
\label{algo}

As its input, the algorithm receives a video of a user watching a
tablet screen as captured from the front camera. The intrinsic (or
extrinsic) camera parameters are unknown. As its output the algorithm
predicts the user's distance (in metric units) from the tablet screen in each video
frame.  There is no expectation that the user will be in a seated
position, or in any fixed position.  It is conceivable that the user
may not be visible while the task is going on since children can be
easily distracted. 

\begin{wrapfigure}{r}{0.2\textwidth}
\includegraphics[width=0.2\textwidth]{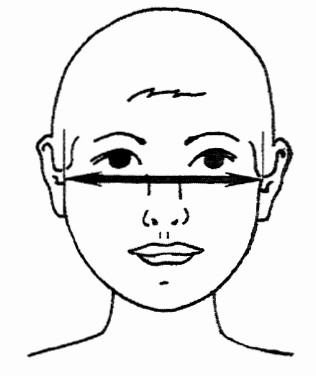}
The length, denoted by the arrowheads, represent the bitragion breadth
(\cite{115-bitragion}, page 188).
\label{bitragion figure}
\end{wrapfigure}
First we detected the face, and then landmarks within the face using a
state-of-the-art deep neural network-based face detector
(FAN)~\cite{114-FAN}.  This detector provides, for each frame in the
video, a reference 3D value for each landmark in a local coordinate
system. However, these 3D coordinates, while valid for each frame, are
incomparable across frames.  Next, a reference frame is therefore
desirable, and we used the frame corresponding to the ``most frontal''\footnote{Details provided in the supplementary material}
face amongst all faces in the video.

The unifying factor across all these frames is, of course, the
camera. Since the 2D landmark for each corresponding 3D landmark is
available (indeed, that's how the 3D is estimated in the first place),
we were able to use these correspondences to find the camera centre.
Once the camera centre is computed, it is relatively straightforward
to compute the metric distance to one of the landmarks, such as the
left or the right eye. The data from this step is used in ablation \hyperlink{ablation}{Scheme 2} described later. 

As communicated by domain experts, some child development studies rely on actual Euclidean distance values. Since from a single camera, we cannot obtain Euclidean (metric)
values, we use the concept of ``bitragion breadth'' -- the width
between the two tragions (cartilaginous notches at the front of the
ear).  Once the head pose is calculated in pixel units, we obtain the
average depth in real units using the scale factor from the bitragion
breadth of children. The benchmark bitragion breadth is taken to be
10.6 cm corresponding to the $50^{th}$ percentile bitragion breadth
for children, both males and females (obtained from
\cite{115-bitragion}, page 454). While the head sizes do vary across children, the variation is insignificant in children in our data set who primarily belong in a narrow age range.

\subsubsection{Algorithm Correctness}
\begin{figure}[!h]
	\centering
	\begin{minipage}[t]{0.47\textwidth}
		\includegraphics[width=\linewidth]{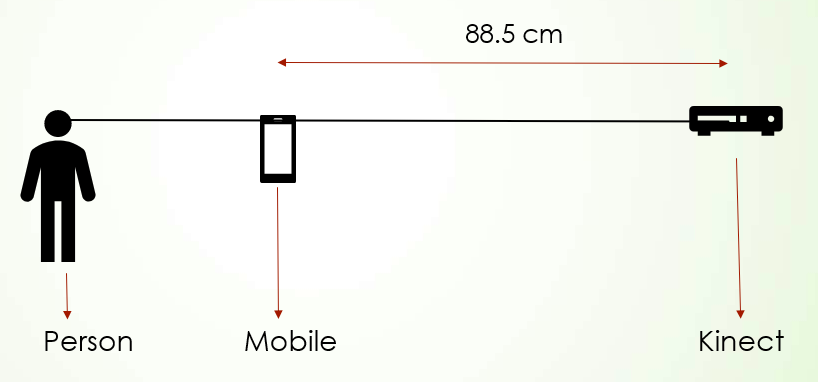}
		\subcaption{Setup} 
		\label{fig:kinect setup}
	\end{minipage} \hfill
	\begin{minipage}[t]{0.47\textwidth}
		\includegraphics[width=\linewidth]{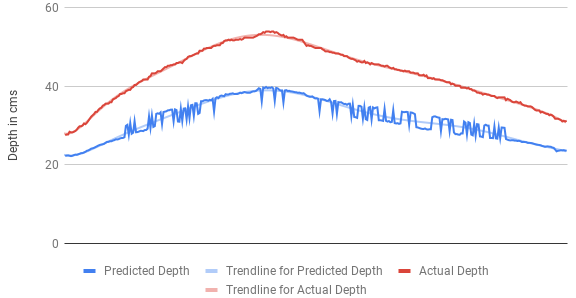}
		\subcaption[Viewer Proximity ASD]{Validation.}
		\label{fig:s1t3}
	\end{minipage}
	\vspace{0.4cm}
	\caption[Proximity Estimation]{(a) Representative depiction of
          the Kinect setup to validate the correctness of our
          algorithm. The actual computation does not use the Kinect.
          (b) Distance in cm vs the frame number. Our results mimic those of the Kinect. The difference in the estimate in (b) is due to the pixel scaling factor used in our approach which assumes a small child's head while the validations were performed on adults.}
\end{figure}

To validate our approach we need reliable ground truth values for viewer distance from the tablet screen. This ground truth can then be used for validating the results output by our proximity estimation algorithm. We followed two different approaches for validation. The first approach relied on manually noting the distances for each frame while the second approach relied on a Kinect V2 based annotation setup. Here, we discuss validation using the Kinect device based on the setup shown in Figure~\ref{fig:kinect setup} (both approaches revealed similar results).  In the setup we have placed the tablet device at a distance of $90$ cm from the Kinect.


\textbf{Calibrating Kinect:} During the validation process we observed
that Kinect prediction in the near camera range was not reliable
(During the wheel task children typically tend to watch the screen at
a range of $20-50$ cm from the tablet device.  Kinect predictions were
first manually verified to be stable and reliable in the 60+ cm range
(hence the $90$ cm distance mentioned above). More details about Kinect
calibration are included in the supplementary material.


\textbf{Validating the algorithm:} 
The output of a sample test run on this validation set up is depicted
in Figure~\ref{fig:s1t3}. In the setup, the viewer is looking at the
mobile screen and is initially 30 cm away from it. The viewer moves
back up to a distance of 50 cm away from the mobile before returning
to the starting point. Note that this translates to a corresponding
distance of 118.5 and 138.5 cm in the setup, respectively (refer
Figure~\ref{fig:kinect setup}).

Although they follow the same trend, as shown in Figure
\ref{fig:s1t3}, the distance predicted by our approach do not match
the ground truth (values predicted by Kinect). This is expected as the
predictions are derived from an assumed bitragion breadth of 10.6 cm
which is typically the value for an average child (the corresponding
size for an average adult would be considerably higher. The validation
tests were performed by adults). The predicted depth values are
filtered (to smoothen out and remove the noise in prediction) before
being used for diagnostic classification.
\begin{figure}[!h]
	\centering
	\begin{minipage}[t]{0.47\textwidth}
	    \centering
		\includegraphics[width=\linewidth]{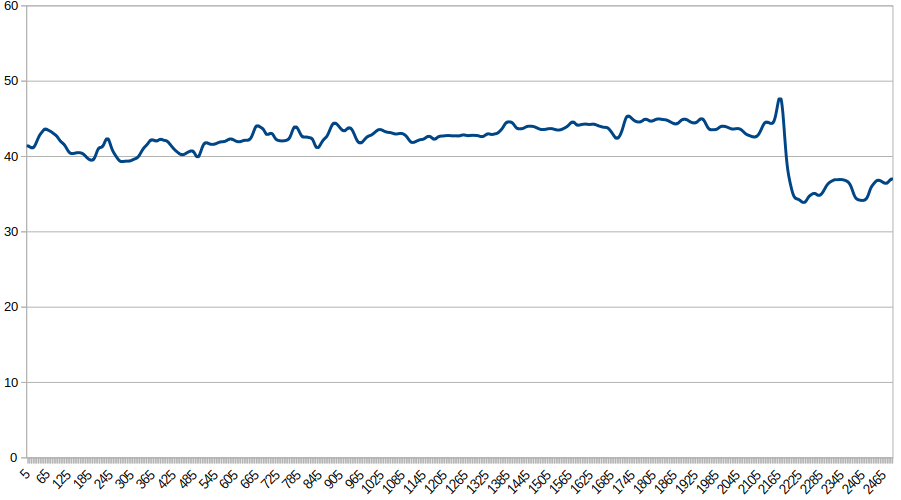}
		\subcaption[Viewer Proximity TD]{Typically Developing category}
		\label{fig:TD proximity}
	\end{minipage}\hfill
	\begin{minipage}[t]{0.47\textwidth}
		\includegraphics[width=\linewidth]{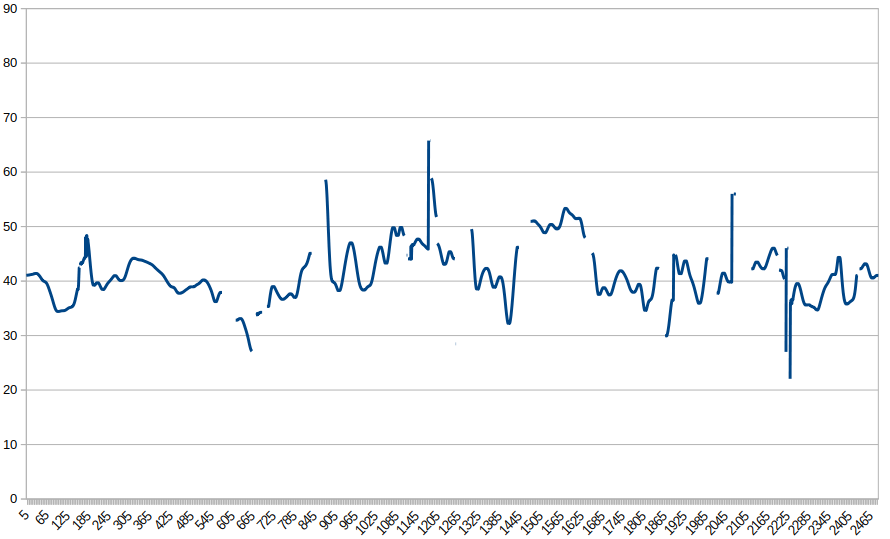}
		\subcaption[Viewer Proximity ASD]{Autism Spectrum Disorder category}
		\label{fig:ASD proximity}
	\end{minipage}
	\vspace{0.4cm}
	\caption[Viewer Proximity]{Distance in cm vs frame number of the video. These figures show the motion of a child's head relative to the tablet screen while performing the wheel task from our viewer proximity estimation algorithm.}
\end{figure}

\subsection{Training}\label{2.3}

\begin{figure*}
\centering
\begin{subfigure}[b]{.45\linewidth}
\includegraphics[width=.9\linewidth]{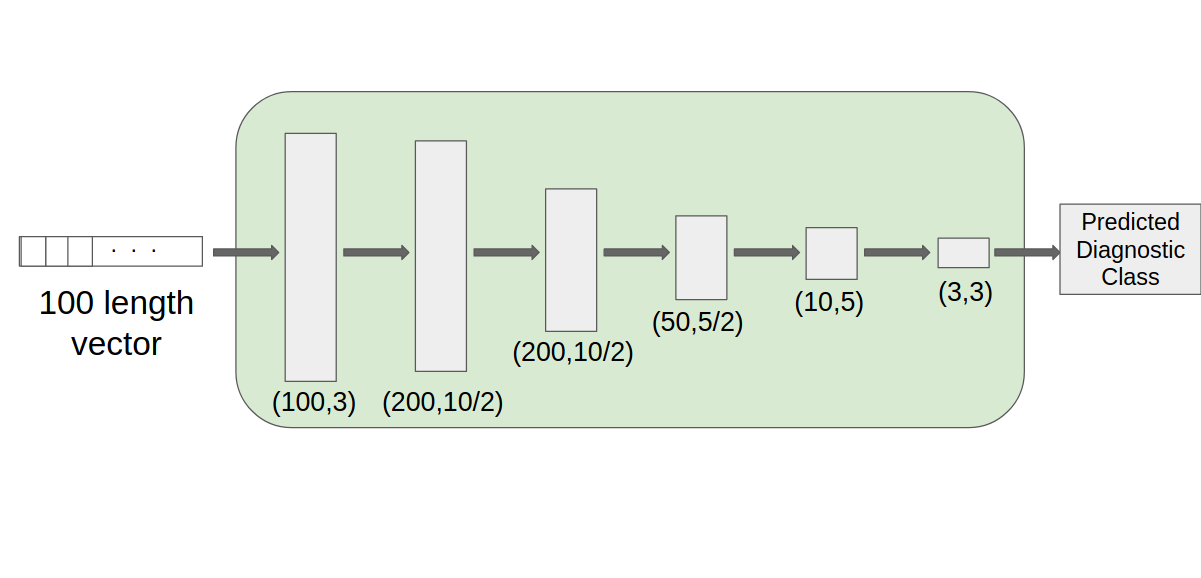}
\caption{Model used in Schemes 1 and 2}\label{fig:Model1}
\end{subfigure}
\begin{subfigure}[b]{.45\linewidth}
\includegraphics[width=.9\linewidth]{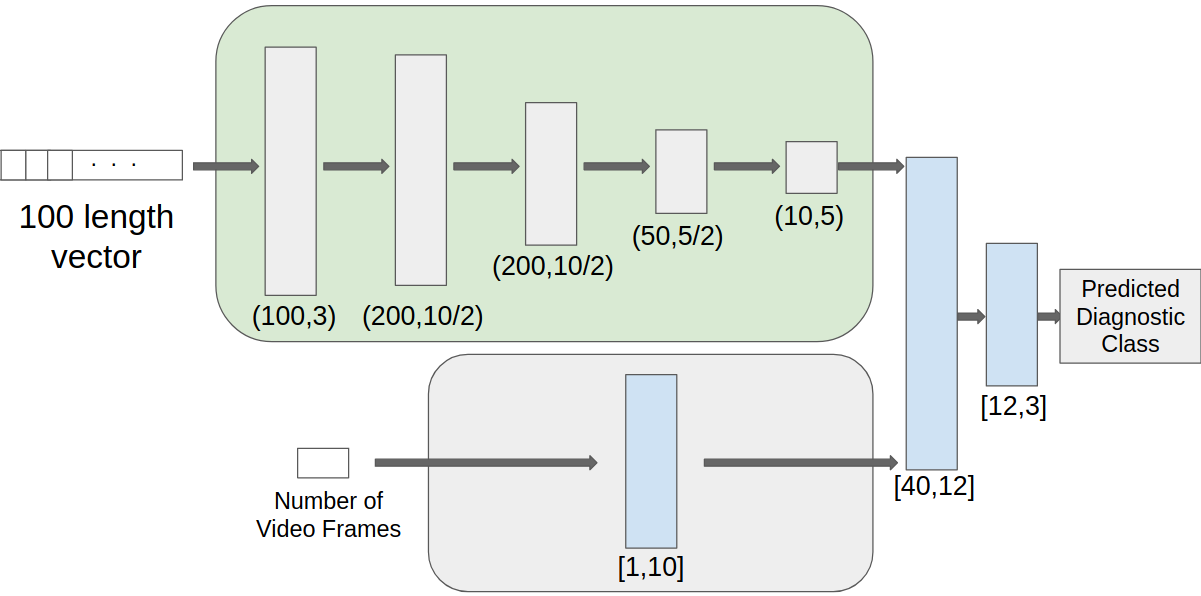}
\caption{Model used in Scheme 3}\label{fig:Model2}
\end{subfigure}
\caption{Network models used across the different schemes. (a) Uses only the fixed length distance vectors as input. Specification provided below each 1D convolution layer as (out channels, kernel size/stride) (b) Additionally uses video duration as an input. Specification of fully connected layers depicted as [in features, out features]. The output from two different channels is concatenated and fed forward to fully connected layers}
\label{fig:Models}
\end{figure*}

In this section, we describe the process of analysing the viewer movement data in order to predict the diagnostic
classification. We utilise the data output by our proximity estimation
algorithm as described in Section~\ref{algo}.

\textbf{Preprocessing:} As is evident from Figure~\ref{fig:TD
  proximity} and Figure~\ref{fig:ASD proximity} the movement data is
now reduced to a vector containing the estimated distance of the child
from the tablet for each frame of the recorded video. The size of the
vector depends on how long the task was undertaken (a stop
button is provided on the screen for the child).  It can range from a few hundred
data points to a maximum of $2400$ data points (recall that videos in the task are displayed for no more than 80 seconds).

Since the whole process is natural and free format, there are frames
which do not depict the child, i.e., the child has moved out of the
screen. These could be a deliberate action on the part of the child
(``disinterest'') or could be a momentary glitch (``the tablet is
swung out'').  One way to handle this is to apply a windowed median
filter followed by Gaussian smoothing. Any other kind of preprocessing, e.g. applying transforms, was not considered as the neural network was expected to learn from the raw data. 

For larger gaps, there may have been some intent, and we wish to seek
to preserve this information. Therefore, if after filtering, the
distance estimate is still unavailable, a marker value (`0') is
substituted with the intention of letting the network learn this
event. (Any resulting string of $0$s in the vector has been replaced
with a single $0$ value.) We observed that this replacement stabilised
the subsequent learning process.

Since the input vector sizes are of variable length, we extracted
contiguous chunks of smaller vectors of size $100$ from each input
vector. This helped in providing a uniform input size as well as, in
essence, boosting the training set size.

The resulting equi-sized vectors are provided as input to a 1D convolutional neural network to predict one of the three diagnostic
classes. The network details are shown in Figure \ref{fig:Models}. Cross entropy loss function was employed to the last layer's output to arrive at the final predictions. 
A data split of $7:1:2$ was used for training, validation, and test sets respectively.

\textbf{Ablation schemes:} To further investigate the impact of various factors on the classification accuracy, we have divided the analysis into three broad schemes as described below:
\begin{itemize}
\item \textbf{Scheme 1}: In this scheme we analyse the movement data as generated by our viewer proximity algorithm (refer Figure~\ref{fig:TD proximity}). The equi-sized chunks of vectors are input as is to the network model shown in Figure~\ref{fig:Model1}.
\item \textbf{\hypertarget{ablation}{Scheme 2}}: In this scheme we normalise the movement data before providing it as input to the network model as shown in Figure~\ref{fig:Model1}. Since the movement data is estimated based on an average head size (derived using the bitragion width of an average child), this scheme mitigates any inadvertent impact of this assumption. The movement data for each child subject is normalised by mean shifting and scaling it down by the respective standard deviation.
\item \textbf{Scheme 3}: As mentioned earlier the output video duration in our data set differ across child subjects. This duration indicates how long the child was actually engaged in the wheel task. In this scheme we analyse the impact of supplementing the input with video duration (expressed as the number of frames in the video). 

To analyse this modified input we have altered our network model. Now, there are two separate inputs provided:
\begin{itemize}
\item vector representing the movement data. This part is unaltered and remains consistent with schemes 1 and 2 described above. The output of the penultimate layer is extracted before final inference and concatenated as shown in Figure~\ref{fig:Model2}
\item number of video frames for the corresponding video. This input is parsed through a fully connected layer and concatenated with the output of the penultimate layer mentioned above
\end{itemize}
The concatenated vector is finally parsed through fully connected layers for making the diagnostic classification prediction.


\end{itemize}
\section{Results}
\label{sec:results}
A 5-fold cross validation approach was utilised for estimating
the network accuracy. Five sets or folds of mutually exclusive data
were created with almost equal representation from subjects of each
diagnostic class. The selection was randomly performed with no manual
considerations. Table~\ref{table:Data Split} lists the data split
for each of the five sets.


\begin{table}[!h]
\begin{center}
\begin{tabular}{|l|c|c|c|c|c|}
\hline
\textbf{Category} & Fold1 & Fold2 & Fold3 & Fold4 & Fold5\\
\hline\hline
TD & 7 & 8 & 7 & 8 & 7         \\
ASD & 8 & 8 & 8 & 8 & 9       \\
ID  & 7 & 8 & 7 & 8 & 7        \\
\hline
Total & 22 & 24 & 22 & 24 & 23 \\
\hline
\end{tabular}
\end{center}
\caption[Fold wise Data Split]{Tabular representation of the cross
  validation process listing the number of children in each diagnostic
  category.} 
\label{table:Data Split}
\end{table}

In each of the five rounds, a different hold out set was chosen from
among the five sets. The hold out set was used for testing the model
which was trained (and validated) on the remaining 4 sets. Throughout
the five rounds a consistent split of $7:1$ was maintained between the training and validation sets.
\newline


Note that a naive classifier, which stubbornly outputs only one class
for any subject, would have achieved a maximum classification accuracy
of 35.65\% ($=41/115$). A random scheme would get one out of 3 right
(33\%). 

\begin{figure}
\centering
\begin{subfigure}[b]{.5\textwidth}
\centering
\includegraphics[width=.75\textwidth]{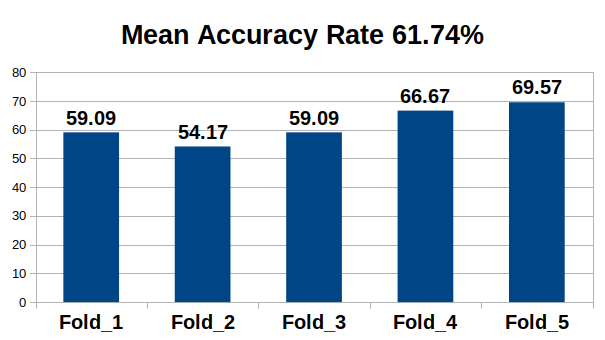}
\caption{Scheme 1}\label{fig:Scheme1a}
\end{subfigure}

\begin{subfigure}[b]{.5\textwidth}
\centering
\includegraphics[width=.75\textwidth]{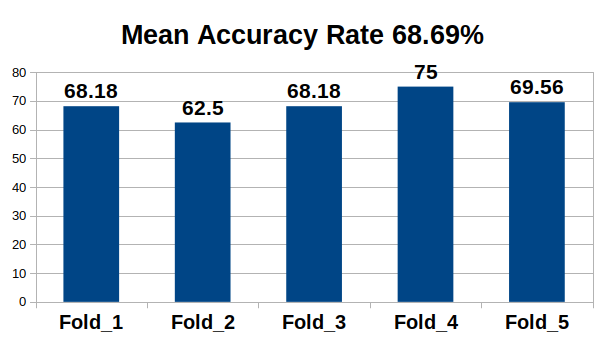}
\caption{Scheme 2}\label{fig:Scheme2a}
\end{subfigure}

\begin{subfigure}[p]{.5\textwidth}
\centering
\includegraphics[width=0.75\textwidth]{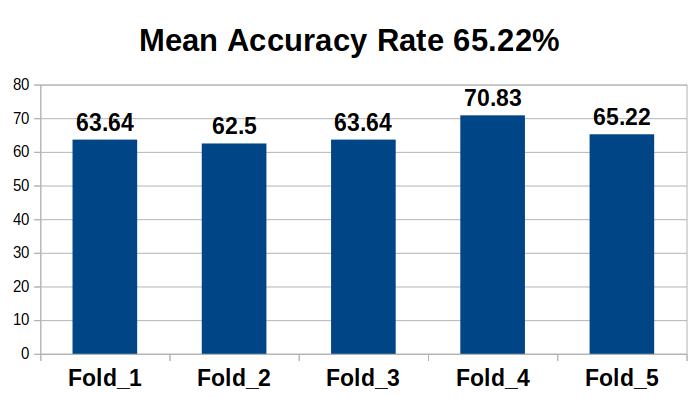}
\caption{Scheme 3}\label{fig:Scheme3a}
\end{subfigure}

\caption{Accuracy results for the different analysis schemes}
\label{fig:AccuracyResults}
\end{figure}

It is to be noted that the network has been
trained for predicting the label for smaller equi-sized input
vectors. The final classification for a subject (child) was arrived at by considering the
modal value of the predictions for all the
constituent smaller vectors for that subject.
The resulting prediction accuracies of the models, for the corresponding schemes, have been noted for each of the five rounds as shown in Figure~\ref{fig:AccuracyResults}. Please note that the average accuracy is arrived at by weighting the round's accuracy by the size of the corresponding hold out set (i.e., number of subjects in the hold out set).  

We observe that Scheme~2 provides better results compared to the other schemes. It can be noted that normalizing the estimated viewer proximity values enhances the overall classification accuracy to  $68.69\%$.
Further, we present the confusion matrix for one of the 5 folds of the experiment for Scheme~2. Here, the columns represent the predicted values for a given class.  
\begin{wrapfigure}{r}{0.3\textwidth}
    \begin{tabular}{|l|c|c|c|}
    \hline
    \textbf{Category} & TD & ASD & ID  \\
    \hline\hline
    TD & \cellcolor[HTML]{C0C0C0}6 & 2   & 0    \\
    ASD & 0   & \cellcolor[HTML]{C0C0C0}8   & 0  \\
    ID & 2    & 2   & \cellcolor[HTML]{C0C0C0}4   \\
    \hline
    \end{tabular}
\end{wrapfigure}
Similar confusion matrices for the remaining folds used for
cross-validation have been included in the supplementary material. 

\section{Conclusion}
Finding behavioural markers for ASD is a challenging proposition.  Nearly all prior work in this direction have only been experimented in a laboratory or a
clinic setup. It is desirable if the hospital comes to the child but such a setup poses interesting challenges. This process becomes even more difficult if the hospital comes to the child in the form of a mobile device that has ``tasks''.  

In this paper, we analysed the ``wheel task'' in which the front
camera of the device is used to record a child watching a ``spinning''
wheel.  To the best of our knowledge, such an analysis has not been
performed in the casual setting of a semi-urban home in LMIC
countries. We used the viewer proximity as a marker for ASD, and we
observed that we could correctly classify a child solely with this
measure with an average accuracy of 68.69\%. Although this may not seem like a strong result when compared to classification tasks in other domains, it is the first such attempt for this task. Such tasks are usually carried out in lab settings, where the analysis is subjective. Further, our results may be seen as a weak classifier which can assist in such analyses. It is also worth noting that this accuracy is significantly higher
than that of a naive classifier.  We are able to achieve this through
a computer vision algorithm that measures viewer proximity in casual
video recordings.
\section{Acknowledgement} The START (Screening Tools for Autism Risk using Technology) consortium members are listed alphabetically: Bhismadev Chakrabarti, Debarati Mukherjee, Gauri Divan,  Georgia Lockwood-Estrin, Indu Dubey, Jayashree Dasgupta, Mark Johnson, Matthew Belmonte, Rahul Bishain, Sharat Chandran, Sheffali Gulati, Supriya Bhavnani, Teodora Gliga, Vikram Patel. This work was funded by a Medical Research Council Global Challenge Research Fund grant to the START consortium (PI: Chakrabarti; Grant ID: MR/P023894/1).
\bibliographystyle{IEEEtran} 
\bibliography{eg2bib.bib}

\begin{thebibliography}{10}
\providecommand{\url}[1]{#1}
\csname url@samestyle\endcsname
\providecommand{\newblock}{\relax}
\providecommand{\bibinfo}[2]{#2}
\providecommand{\BIBentrySTDinterwordspacing}{\spaceskip=0pt\relax}
\providecommand{\BIBentryALTinterwordstretchfactor}{4}
\providecommand{\BIBentryALTinterwordspacing}{\spaceskip=\fontdimen2\font plus
\BIBentryALTinterwordstretchfactor\fontdimen3\font minus
  \fontdimen4\font\relax}
\providecommand{\BIBforeignlanguage}[2]{{%
\expandafter\ifx\csname l@#1\endcsname\relax
\typeout{** WARNING: IEEEtran.bst: No hyphenation pattern has been}%
\typeout{** loaded for the language `#1'. Using the pattern for}%
\typeout{** the default language instead.}%
\else
\language=\csname l@#1\endcsname
\fi
#2}}
\providecommand{\BIBdecl}{\relax}
\BIBdecl

\bibitem{dsm5}
\emph{Diagnostic and statistical manual of mental disorders}, 5th~ed., American
  Psychiatric Association, 2013.

\bibitem{egger2018}
H.~L. Egger, G.~Dawson, J.~Hashemi, K.~L. Carpenter, S.~Espinosa, K.~Campbell,
  S.~Brotkin, J.~Schaich-Borg, Q.~Qiu, M.~Tepper, J.~P. Baker, R.~A.
  Bloomfield, and G.~Sapiro, ``Automatic emotion and attention analysis of
  young children at home: a {R}esearch{K}it autism feasibility study,''
  \emph{NPJ {D}igital {M}edicine}, vol.~1, no.~1, pp. 1--10, 2018.

\bibitem{intraface}
X.~Xiong and F.~De~la Torre, ``Supervised descent method and its applications
  to face alignment,'' in \emph{CVPR}, 2013, pp. 532--539.

\bibitem{Hashemi}
J.~Hashemi, K.~Campbell, K.~L. Carpenter, A.~Harris, Q.~Qui, M.~Tepper,
  S.~Espinosa1, B.~J. S., S.~Marsan, R.~Calderbank, J.~Baker, H.~L. Egger,
  G.~Dawson, and G.~Sapiro, ``A scalable app for measuring autism risk
  behaviours in young children,'' in \emph{MOBIHEALTH}, 2015.

\bibitem{bhisma}
S.~Baron-Cohen, E.~Ashwin, C.~Ashwin, T.~Tavassoli, and B.~Chakrabarti,
  ``Talent in autism: hyper-systemizing, hyper-attention to detail and sensory
  hypersensitivity,'' \emph{Philosophical Transactions of the Royal Society B:
  Biological Sciences}, vol. 364, no. 1522, pp. 1377--1383, 2009.

\bibitem{tavassoli}
T.~Tavassoli, K.~Bellesheim, P.~M. Siper, A.~T. Wang, D.~Halpern,
  M.~Gorenstein, and J.~D. Buxbaum, ``Measuring sensory reactivity in autism
  spectrum disorder: application and simplification of a clinician-administered
  sensory observation scale.'' \emph{Journal of Autism and Developmental
  Disorders}, vol.~46, no.~1, pp. 287--293, 2016.

\bibitem{112-NIH}
S.~Freeman and C.~Kasari, ``Parent--child interactions in autism:
  Characteristics of play,'' \emph{Autism}, vol.~17, no.~2, pp. 147--161, 2013.

\bibitem{103-ADOS}
J.~M. Guercio and A.~D. Hahs, ``Applied behavior analysis and the autism
  diagnostic observation schedule ({ADOS}): A symbiotic relationship for
  advancements in services for individuals with autism spectrum disorders
  ({ASDs}),'' \emph{Behavior analysis in practice}, vol.~8, no.~1, pp. 62--65,
  2015.

\bibitem{104-DCMA}
A.~Pickles, A.~Le~Couteur, K.~Leadbitter, E.~Salomone, R.~Cole-Fletcher,
  H.~Tobin, I.~Gammer, J.~Lowry, G.~Vamvakas, S.~Byford, and C.~Alfred,
  ``Parent-mediated social communication therapy for young children with autism
  ({PACT}): long-term follow-up of a randomised controlled trial,'' \emph{The
  Lancet}, vol. 388, no. 10059, pp. 2501--2509, 2016.

\bibitem{6619282}
J.~Rehg, G.~Abowd, A.~Rozga, M.~Romero, M.~Clements, S.~Sclaroff, I.~Essa,
  O.~Ousley, Y.~Li, C.~Kim, and H.~Rao, ``Decoding children's social
  behavior,'' in \emph{CVPR}, 2013, pp. 3414--3421.

\bibitem{110-Marinou}
E.~Marinoiu, M.~Zanfir, V.~Olaru, and C.~Sminchisescu, ``3{D} human sensing,
  action and emotion recognition in robot assisted therapy of children with
  autism,'' in \emph{CVPR}, 2018, pp. 2158--2167.

\bibitem{Martin2018}
K.~B. Martin, Z.~Hammal, G.~Ren, J.~F. Cohn, J.~Cassell, M.~Ogihara, J.~C.
  Britton, A.~Gutierrez, and D.~S. Messinger, ``Objective measurement of head
  movement differences in children with and without autism spectrum disorder,''
  \emph{Molecular Autism}, vol.~9, no.~1, p.~14, 2018.

\bibitem{Ogihara_2019_CVPR_Workshops}
M.~Ogihara, Z.~Hammal, K.~B. Martin, J.~F. Cohn, J.~Cassell, G.~Ren, and D.~S.
  Messinger, ``Categorical timeline allocation and alignment for diagnostic
  head movement tracking feature analysis,'' in \emph{CVPR Workshops}, 2019,
  pp. 43--51.

\bibitem{115-bitragion}
\BIBentryALTinterwordspacing
R.~G. Snyder, L.~W. Schneider, L.~O. Clyde, M.~R. Herbert, D.~H. Golomb, and
  M.~A. Schork, ``Anthropometry of infants, children, and youths to age 18 for
  product safety design. final report.'' 1977, last accessed: Apr 2020.
  [Online]. Available:
  \url{https://math.nist.gov/~SRessler/anthrokids/child77lnk.pdf}
\BIBentrySTDinterwordspacing

\bibitem{114-FAN}
A.~Bulat and G.~Tzimiropoulos, ``How far are we from solving the 2{D} \& 3{D}
  face alignment problem?(and a dataset of 230,000 3{D} facial landmarks),'' in
  \emph{ICCV}, 2017, pp. 1021--1030.

\end{thebibliography}


\end{document}


\graphicspath{{images/}}
\section{Supplementary Material}
\subsection{Sample Videos}
Please refer to the attached video files for a demonstration
\subsubsection{Sample wheel task video}
This attached video \attachfile[mimetype=video/webm,icon=Paperclip,appearance=false]{sample.webm} provides an example of the
video of a child performing the wheel test in a home setting. Please
note that the video has been modified to protect the privacy of the
participant. (For instance, the yellow box is added to detect and
subsequently blur the child's face).
\subsubsection{START A field data collection app for autism screening}
The wheel test is one of several assessments which are used for
screening. We collected data on various other tests by taking a tablet
directly to households. Like the wheel test, these assessments too are
part of an app which is installed on the tablet.  For context, we
provide this attached video \attachfile[mimetype=video/webm,icon=Paperclip,appearance=false]{bringing.webm}
\subsection{Kinect Calibration}
\begin{figure}[h!]
\centering
\includegraphics[width=0.48\textwidth]{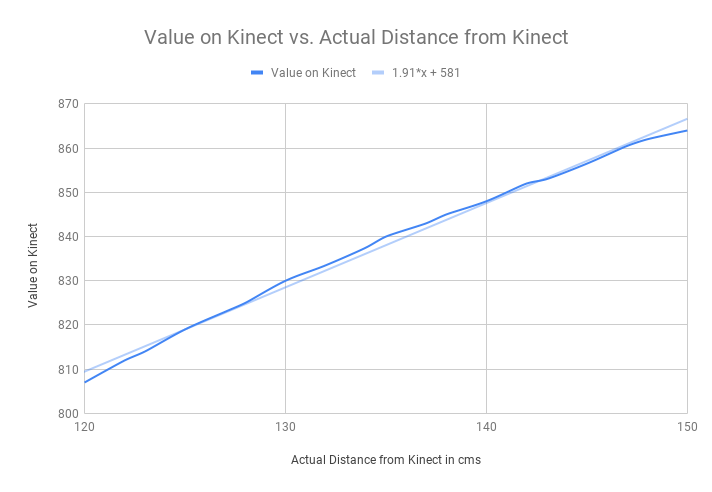}
\caption{Correlation between Kinect predicted depth values and the actual distance in cm. The Y-axis represents the depth value as predicted by Kinect. The corresponding ground truth observed distance is plotted on the X-axis. This correlation is used to convert the Kinect predictions to the actual distance in centimetres.}
\label{fig:kinect-calib}
\end{figure}

Kinect depth prediction values were not in centimeter units. We have
arrived at the correlation between the Kinect predicted depth values
to the actual distance in cm, as shown in
Figure~\ref{fig:kinect-calib}. To arrive at this relation both the
Kinect device and a plane object were kept on a level surface. The
object was then kept at various distances in the 90-120 cm range from
the Kinect. The predicted Kinect depth is plotted against the actual
object distance to arrive at the approximation given by the following
formula $y = 1.91x + 581$, where
$y$ is the depth of the object as predicted by Kinect, and $x$ is the
ground truth observed distance of the object.

\subsection{Deriving most frontal face}
In this section we provide the steps to determine the frame with the most ‘frontal’ face in a video. The intention is to capture the image frame in which the user’s face tends to be most upright and parallel to the image plane. 

The following steps are performed for arriving at the most frontal face:

\textbf{Input}: 68 facial landmarks as shown in Figure~\ref{fig:landmarks} for each frame 

\textbf{Output}: Frame number with the most frontal face

\textbf{Calculation steps}:
\begin{enumerate}
    \item For each image frame in the video:
        
    \begin{enumerate}
        \item Mean shift the landmarks (using mean of the landmarks in the current frame)
        \item Scale each landmark down by its corresponding magnitude
        \item \label{depth} Obtain the (absolute) depth difference for opposite landmark point pairs on the jaw-line (landmark point pairs – 1:17, 2:16, 3:15, ...8:10). Select the maximum ‘depth difference’ for this image frame from these values 
    
    \begin{figure}
    \centering
    
    \begin{subfigure}[b]{.5\textwidth}
    \centering
    \includegraphics[width=.65\textwidth]{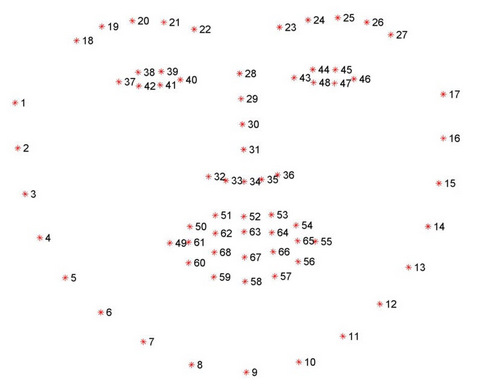}
    \caption{2D facial landmarks}\label{fig:Scheme1a}
    \end{subfigure}
    
    \begin{subfigure}[b]{.5\textwidth}
    \centering
    \includegraphics[width=0.9\textwidth]{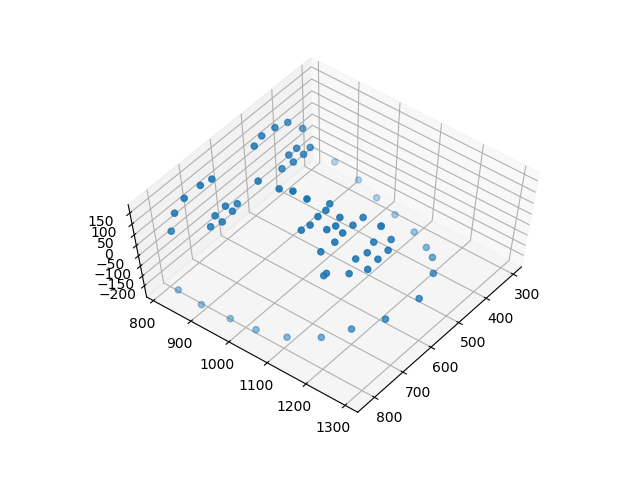}
    \caption{3D facial landmarks}\label{fig:Scheme3a}
    \end{subfigure}
    
    \caption{Plots of (a) 2D and (b) 3D facial landmarks used to arrive at the most frontal face}
    \label{fig:landmarks}
    \end{figure}
    \end{enumerate}    
            
    \item From among the frames having $\pm4$ degree ‘eyeline orientation’, select the min(max(depth difference)). i.e. Note the frame with with minimum of the maximum depth difference calculated in step~\ref{depth} above.

    For calculation of ‘eyeline orientation’ the steps are:
        \begin{enumerate}
            \item Obtain centroid of eyes (centroid of landmark points 37-42 and 43-48)
            \item Obtain eyeline orientation as the slope of the line joining the two eye centroids
        \end{enumerate}
    \end{enumerate}
\subsection{Confusion Matrices}

We present the confusion matrix for each of the five folds below (see
main paper for description) for Scheme~2 here. Note that the precision and recall value
for each fold is indicated in the last row and column, respectively of
the corresponding fold. As before, columns represent the predicted
values for a given class.

The overall precision and recall values have been depicted in Table~\ref{table:AvgCM}

\begin{table}[H]
\parbox{.4\textwidth}{
\begin{tabular}{|l|c|c|c|c|}
\hline
\textbf{Category} & TD & ASD & ID & Recall \\
\hline\hline
TD & \cellcolor[HTML]{C0C0C0}5&0&2& 0.71\\
ASD & 0& \cellcolor[HTML]{C0C0C0}7&2& 0.78\\
ID & 0&3& \cellcolor[HTML]{C0C0C0}4& 0.57\\
\hline Precision & 1 & 0.7& 0.5& \\

\hline
\end{tabular}
\caption[Fold 1 confusion matrix]{Fold 1 confusion matrix} 
\label{table:Fold1 CM}
}
\parbox{.4\textwidth}{
\begin{tabular}{|l|c|c|c|c|}
\hline
\textbf{Category} & TD & ASD & ID & Recall \\
\hline\hline
TD & \cellcolor[HTML]{C0C0C0}6&2&0& 0.75\\
ASD & 0& \cellcolor[HTML]{C0C0C0}8&0& 1\\
ID & 2&2& \cellcolor[HTML]{C0C0C0}4& 0.5\\
\hline Precision & 0.75 & 0.67& 1& \\

\hline
\end{tabular}
\caption[Fold 2 confusion matrix]{Fold 2 confusion matrix} 
\label{table:Fold2 CM}
}
\parbox{.4\textwidth}{
\begin{tabular}{|l|c|c|c|c|}
\hline
\textbf{Category} & TD & ASD & ID & Recall \\
\hline\hline
TD & \cellcolor[HTML]{C0C0C0}5&1&1& 0.71\\
ASD & 3& \cellcolor[HTML]{C0C0C0}5&0& 0.63\\
ID & 1&1& \cellcolor[HTML]{C0C0C0}5& 0.71\\
\hline Precision & 0.56 & 0.71& 0.83& \\

\hline
\end{tabular}

\caption[Fold 3 confusion matrix]{Fold 3 confusion matrix} 
\label{table:Fold3 CM}
}
\parbox{.4\textwidth}{
\begin{tabular}{|l|c|c|c|c|}
\hline
\textbf{Category} & TD & ASD & ID & Recall \\
\hline\hline
TD & \cellcolor[HTML]{C0C0C0}4&2&2& 0.5\\
ASD & 0& \cellcolor[HTML]{C0C0C0}7&1& 0.88\\
ID & 0&4& \cellcolor[HTML]{C0C0C0}4& 0.5\\
\hline Precision & 1 & 0.54& 0.57& \\

\hline
\end{tabular}

\caption[Fold 4 confusion matrix]{Fold 4 confusion matrix} 
\label{table:Fold4 CM}
}
\parbox{.4\textwidth}{
\begin{tabular}{|l|c|c|c|c|}
\hline
\textbf{Category} & TD & ASD & ID & Recall \\
\hline\hline
TD & \cellcolor[HTML]{C0C0C0}6&1&0& 0.86\\
ASD & 0& \cellcolor[HTML]{C0C0C0}7&1& 0.88\\
ID & 1&4& \cellcolor[HTML]{C0C0C0}2& 0.29\\
\hline Precision & 0.86 & 0.58& 0.67& \\

\hline
\end{tabular}

\caption[Fold 5 confusion matrix]{Fold 5 confusion matrix} 
\label{table:Fold5 CM}
}

\parbox{.4\textwidth}{
\begin{tabular}{|l|c|c|c|c|}
\hline
\textbf{Category} & TD & ASD & ID & Average Recall \\
\hline\hline
TD & \cellcolor[HTML]{C0C0C0}26&6&5& 0.7\\
ASD & 3& \cellcolor[HTML]{C0C0C0}34&4& 0.83\\
ID & 4&14& \cellcolor[HTML]{C0C0C0}19& 0.51\\
\hline Average Precision & 0.79 & 0.63& 0.68& \\
\hline
\end{tabular}

\caption[Mean Precision and Recall]{\textbf{Overall Precision and Recall}} 
\label{table:AvgCM}
}

\end{table}

\subsection{Mean reciprocal rank (MRR)}

We report an average MRR of 0.83 for the five folds for Scheme~2. The MRR values for individual folds are as shown below in Table~\ref{table:MRR}

\begin{table}[H]
\centering
\parbox{.4\textwidth}{
\begin{tabular}{|l|c|c|c|c|c|c|}
\hline
\textbf{Fold No.} & 1 & 2 & 3 & 4 & 5 & \textbf{Overall} \\
\hline\hline
\textbf{MRR} & 0.826 & 0.792 & 0.818 & 0.854 & 0.833 & 0.825 \\
\hline
\end{tabular}
\caption[MRR]{Mean Reciprocal Rank for Scheme 2} 
\label{table:MRR}
}
\end{table}